\documentclass[prl,twocolumn,amsmath,nofootinbib,showpacs]{revtex4}
\usepackage{bm,amssymb}
\DeclareMathAlphabet{\varmathbb}{U}{bbold}{m}{n}

\newcommand{\bfone}{{\bm 1}}

\newcommand{\bfP}{\bm{P}}

\newcommand{\bfepsilon}{{\bm \epsilon}}
\newcommand{\eqn}[1]{Eq.\ (\ref{Eqn:#1})}
\newcommand{\eqns}[1]{Eqs.\ (\ref{Eqn:#1})}
\newcommand{\eqnref}[1]{(\ref{Eqn:#1})}

\newcommand{\opp}[2]{\ensuremath{\frac{d #1}{d #2}}}

\begin{document}
\title{Steady-state relaxation and the first passage time distribution of the generalized master equation}
\author{David Shalloway}
\affiliation{Biophysics Program, Dept. of Molecular Biology and Genetics, Cornell University, Ithaca, New York  14853, USA}
\homepage{http://www.mbg.cornell.edu/shalloway/shalloway.html}
\email{dis2@cornell.edu}
\thanks{I am indebted to Tony Faradjian and Ron Elber for bringing this problem to my attention and for many helpful discussions.}
\begin{abstract}
In principle, the generalized master equation can be used to efficiently compute the macroscopic first passage time (FPT) distribution of a complex stochastic system from short-term microscopic simulation data. However, computing its transition function matrix, $\Gamma(\tau)$, from such data can be practically difficult or impossible. We solve this problem by showing that the FPT moment generating function is a simple function of the (easily computable) Laplace transform of the local FPT distribution matrix. Physical insight into this relationship is obtained by analyzing the process of steady-state relaxation.
\end{abstract}
\maketitle

Computing the macroscopic transition rates and first passage time
(FPT) distributions of complex systems (e.g., proteins) by
computational simulations using microscopic equations of motion
can be expensive, if not impossible, particularly when
the systems are large and/or when transitions are rare. In
principle, computation can be simplified and comprehension can be
enhanced by coarse-graining the microscopic equations to a
macroscopic \emph{generalized master equation} \cite{Kenkre:74}
\begin{equation}
\label{Eqn:generalized} 
\opp{\bfP(t)}{t}  = \delta(t) \bfP(0) -\int_0^\infty \Gamma(\tau) \cdot \bfP(t-\tau) d\tau \;,
\end{equation}
with  $\bfP(-\infty) = 0$.  This describes the time-evolution of $\bfP(t)$, the ensemble
occupation number $N$-vector defined over states $s$, each
corresponding to a subregion of the microscopic phase space, after
injection of systems at time $t=0$. $\Gamma(\tau)$ is the $N \times N$ matrix of transition functions,
which includes memory effects.  Corresponding to conservation of probability and to causality, 
\begin{equation}
\label{Eqn:conservation}
 \bfone \cdot \Gamma(\tau) =  0 \;, \quad
\Gamma_{s,s'}(\tau) \le 0 \quad (s \ne s')\;,
\end{equation}
(where $\bfone$ is the $N$-vector with all components equal to 1).

The task is to determine the rate, or more generally 
the FPT distribution, of transitions from an initial state $i$ to a
final state $f$.  So as to examine \emph{first} passage times, we
make $f$ an absorptive state:
\begin{equation}
\label{Eqn:irreversible}
\Gamma (\tau) \cdot \hat{\bfepsilon}_f = 0  \;,
\end{equation}
where $\hat{\bfepsilon}_s$ denotes the basis vector which has component $s$ equal to 1 and all
other components 0. Then the FPT distribution is 
\begin{equation}
\label{Eqn:varphi} 
\varphi(t) = dP_f(t)/dt \quad \mbox{for }
\bfP(0) = \hat{\bfepsilon}_i \;,
\end{equation}
the mean FPT (MFPT) is the first moment $\langle \langle  \tau \varphi \rangle \rangle$, where
\[
\langle\langle \tau^k \varphi \rangle \rangle \equiv \int_0^\infty \tau^k \varphi(\tau) \, d\tau  \;,
\]
and the transition rate is $\langle \langle \tau \varphi \rangle \rangle^{-1}$.  
We assume that $f$ is the only absorptive state and that the system is ergodic, so 
$\bfP(\infty) = \hat{\bfepsilon}_f$. Therefore,
\begin{equation}
\label{Eqn:normalization}
\int_0^\infty \varphi(\tau) d \tau = P_f(\infty) = 1 \;.
\end{equation}

Defining the Laplace transform of $g(\tau)$ as $\tilde{g}(u) \equiv \int_0^\infty e^{-u \tau} g(\tau) \, d \tau \;,$
the FPT moment generating function is 
\[
\sum_{k=0}^\infty \frac{\langle\langle \tau^k \varphi \rangle
\rangle \alpha^k}{k!}= \tilde{\varphi}(-\alpha) \;. 
\]
We assume (as is true in most cases of interest) that $\varphi(\tau)$ decays faster than $e^{- \alpha_{\rm max} \tau}$ as $\tau \to \infty$ for some positive $\alpha_{\rm max}$, so that $\tilde{\varphi}(-\alpha)$ is analytic in a neighborhood  about 0 and can be differentiated to yield all moments. (This assumption is not essential, but simplifies the discussion.  If it is not true, the discussion will still be valid for the finite moments.) 

In some cases, $\Gamma(\tau)$ can be defined from first
principles. However, for complicated systems such as proteins
it must be computed from microscopic simulations. In
such cases relatively short (compared to the MFPT) molecular or Langevin dynamics simulations can be used to
determine $-K_{s,s'}(\tau) \; (s \ne s')$, the \emph{local FPT distribution}. This is the probability density that, after arriving at state $s'$, a system waits for an interval $\tau$ before first leaving and that it goes to $s$.  The diagonal elements of the $N \times N$ matrix $K(\tau)$ are defined by $K_{s,s}(\tau) \equiv   -\sum_{s \ne s'} K_{s,s'}(\tau)$, so like $\Gamma(\tau)$, $K(\tau)$ satisfies
\[
\bfone \cdot K(\tau) =  0 \;, \quad K_{s,s'}(\tau) \le 0 \quad (s \ne s'), \quad K(\tau) \cdot \bfepsilon_f  =  0 \;,
\]
and also, by its definition and the assumption that $f$ is the only absorptive state, satisfies
\[
\int_0^\infty K_{s,s}(\tau) \, d \tau =  1 \quad (s\ne f) \;.
\]
The Laplace transforms of $\Gamma(\tau)$ and $K(\tau)$ are related by 
\begin{subequations}
\begin{eqnarray}
\label{Eqn:Laplace} 
\widetilde{\Gamma}(u) & = & u \widetilde{K}(u)\cdot \{I- \mbox{Diag}[\widetilde{K}(u)]\}^{-1} \\
\widetilde{K}(u) & = & \widetilde{\Gamma}(u) \cdot \{ uI + \mbox{Diag}[\widetilde{\Gamma}(u)]\}^{-1} \,,
\end{eqnarray}
\end{subequations}
where $I$ is the identity matrix and $\mbox{Diag}[A]$ denotes the matrix of diagonal elements of $A$ \cite{Kenkre:74}. However, even though \eqn{Laplace} can be used to determine $\widetilde{\Gamma}(u)$ from $\widetilde{K}(u)$, the inverse Laplace transform required to determine $\Gamma(\tau)$ can be difficult, if not impossible, to compute. This precludes the computation of $\varphi(t)$ by integration of \eqn{generalized} when only $K(\tau)$ is known.

Faradjian and Elber \cite{Faradjian:04} have recently provided a solution to this problem by showing that \eqn{generalized} can be reformulated using  either a coupled set of \emph{QK equations} or a single set of (pseudo-)Markovian \emph{PJ equations} that allow $d\bfP(t)/dt$ to be integrated in $t$ using only $K(\tau)$, not $\Gamma(\tau)$.    However, this procedure is inefficient since it requires modelling the full functional dependence of $K(\tau)$, even though most of the relevant information is contained within only a few of its lower moments. 

Here we discuss how to compute the FPT moments directly from the moments of $\Gamma(\tau)$ or $K(\tau)$ without need for an alternative formulation and at reduced computational cost. In addition, we show how physical insight into the relationships between the moments emerges from analysis of the physical process of   \emph{steady-state relaxation}.

\emph{Results}---The Laplace transform of \eqn{generalized} with $\bfP(0)= \hat{\bfepsilon}_i$ is 
\[
u \widetilde{P}(u) = \hat{\bfepsilon}_i - \widetilde{\Gamma}(0) \cdot \widetilde{P}(u)
\]
with solution $\widetilde{P}(u) = [uI + \widetilde{\Gamma}(u)]^{-1} \cdot \hat{\bfepsilon}_i$, where $I$ is the identity matrix. Combining this with \eqns{varphi} and \eqnref{Laplace} gives expressions for the FPT moment generating function in terms of either $\widetilde{\Gamma}(u)$ or $\widetilde{K}(u)$:
\begin{subequations}
\label{Eqn:relationships}
\begin{eqnarray}
\label{Eqn:Gamma_Laplace_relationship}
\tilde{\varphi}(-\alpha) & = & - \alpha \hat{\bfepsilon}_f \cdot [\widetilde{\Gamma}(-\alpha) - \alpha I]^{-1} \cdot \hat{\bfepsilon}_i \\
\label{Eqn:K_Laplace_relationship}
& = & \hat{\bfepsilon}_f \cdot [I + \widetilde{\overline{K}}(-\alpha)]^{-1} \cdot \hat{\bfepsilon}_i \;,
\end{eqnarray}
\end{subequations}
where $\overline{A}$ denotes the matrix of off-diagonal elements of $A$ \cite{Elber_note}.  [The singularity in \eqn{Gamma_Laplace_relationship} at $\alpha=0$ is removable.]

To gain insight into the physical significance of these relationships, we replace \eqn{generalized} with the equation for steady-state relaxation---the situation in which systems are continuously injected into the initial state at an exponentially decreasing rate $\exp(-\alpha t)$ beginning at $t=0$. \eqn{generalized} becomes 
\begin{equation}
\label{Eqn:generalized2}
 \opp{\bfP(t)}{t}  = e^{-\alpha t} \theta(t) \hat{\bfepsilon}_i -\int_0^\infty \Gamma(\tau) \cdot \bfP(t-\tau) d\tau \;,
\end{equation}
where $\theta(t)$ is the Heaviside step function.  If $\alpha=0$, as $t \to \infty$ \eqn{generalized2} describes steady-state flow, and we expect that the occupation numbers in all the intermediate states, $P_s(t)$ $(s \ne f)$, will approach a constant steady-state solution. When $\alpha >0$ we expect that the intermediate state occupation numbers will will decay exponentially as $t \to \infty$.  The key point is that the asymptotic steady-state solution is easy to compute and determines the FPT generating function exactly.  

Since \eqns{generalized} and \eqnref{generalized2} are linear, the systems injected at different times will transition to $f$ independently, so $d P_f/dt$ during steady-state relaxation is obtained simply by integrating \eqn{varphi} over the incoming flux:
\begin{equation} 
\label{Eqn:accumulation}
\opp{P_f(t)}{t} = \int_0^t e^{-\alpha (t- \tau)}  \varphi(\tau) \, d \tau   \;.
\end{equation}
That is, $\varphi(\tau)$ determines the fraction of the flux introduced at time $t-\tau$, $e^{-\alpha(t-\tau)}$, that arrives at $f$ at time $t$. As long as $t$ is larger than the support of $\varphi(\tau)$, we can extend the upper limit of the integral to $\infty$, and \eqn{accumulation} can be rewritten as:
\begin{equation}  
\label{Eqn:generating_function}
\hspace{-1em} \tilde{\varphi}(-\alpha) = \int_0^\infty \varphi(\tau) e^{\alpha \tau} \, d \tau  =  e^{\alpha t}  \opp{P_f(t)}{t} \quad (t \gg 0)\;.
\end{equation}
Thus, solving \eqn{generalized2} for $\bfP(t)$ when $t \gg 0$ determines the FPT generating function. 

We must take care because $\Gamma(\tau)$ is singular: \eqns{conservation} and \eqnref{irreversible} imply that it has left null-vector $\bfone$ and corresponding right null-vector $\hat{\bfepsilon}_f$.  Thus, it is convenient to decompose \eqn{generalized2} by projecting it into the null-space and its bi-orthogonal space using the idempotent, asymmetric projection operators $\mathcal{P}$ and $\mathcal{Q}$:
\[
\mathcal{P}  = \hat{\bfepsilon}_f \otimes \bfone , \quad \mathcal{Q} = I - \mathcal{P} \;.
\]
and expanding 
\begin{equation}
\label{Eqn:P_decomposition}
\bfP(t) = \hat{\bfepsilon}_f P_{\rm tot}(t) + \bar{\bfP}(t) \;,
\end{equation}
where $P_{\rm tot}(t) \equiv \bfone \cdot \bfP(t)$ and $\bar{\bfP}(t) \equiv \mathcal{Q} \cdot \bfP(t)$. Substituting this into \eqn{generalized2} yields the two independent equations
\begin{eqnarray*}
\opp{P_{\rm tot}(t)}{t} & = & e^{-\alpha t} \theta(t) \\
\opp{\bar{\bfP}(t)}{t} & = & 
 e^{-\alpha t} \theta(t) (\hat{\bfepsilon}_i - \hat{\bfepsilon}_f) - \int_0^\infty \!\! \Gamma(\tau) \cdot \bar{\bfP}(t-\tau) \, d \tau ,
\end{eqnarray*}
The first equation [along with boundary condition $\bfP(-\infty)=0$] implies that
\begin{equation}
\label{Eqn:Ptot_solution}
P_{\rm tot}(t) = \frac{1-e^{-\alpha t}}{\alpha} \theta(t) \;.
\end{equation}
We solve the second equation in the asymptotic regime by testing the guess that $\bar{\bfP}(t) = \bar{\bfP} \exp(-\alpha t)$ $(t \gg 0)$, where $\bfone \cdot \bar{\bfP} = 0$. Substituting this in, factoring out $\exp(-\alpha t)$, and taking the limit $t \to \infty$ gives 
\begin{subequations}
\label{Eqn:Pbar_solutions}
\begin{eqnarray}
\label{Eqn:Pbar_solutiona}
-\alpha \bar{\bfP} & = & (\hat{\bfepsilon}_i - \hat{\bfepsilon}_f) - \widetilde{\Gamma}(-\alpha) \cdot \bar{\bfP} \\
\label{Eqn:Pbar_solutionb}
\Rightarrow \bar{\bfP} & = & \mathcal{Q} \cdot [\widetilde{\Gamma}(-\alpha) - \alpha I]^{-1} \cdot (\hat{\bfepsilon}_i - \hat{\bfepsilon}_f) \;.
\end{eqnarray}
\end{subequations}

Combining Eqns.\ (\ref{Eqn:generating_function}--\ref{Eqn:Pbar_solutions}) with \eqn{Laplace} gives the generating function as 
\begin{equation}
\label{Eqn:relationship2}
\tilde{\varphi}(-\alpha)   =  1- \alpha \hat{\bfepsilon}_f \cdot  \mathcal{Q} \cdot [\widetilde{\Gamma}(-\alpha) - \alpha I]^{-1} \cdot \mathcal{Q} \cdot \hat{\bfepsilon}_i \;,
\end{equation}
which is equivalent to \eqns{relationships}. In this form it is evident that the zeroth moment is always $\tilde{\varphi}(0)=1$, consistent with \eqn{normalization}.  The higher moments are obtained by expanding in powers of $\alpha$ about $\alpha = 0$ with care for the degeneracy of $\widetilde{\Gamma}(0)$. For example, the first moment in terms of $\widetilde{\Gamma}$ is 
\begin{equation}
\label{Eqn:MFPT}
\langle\langle \tau \varphi \rangle \rangle  =  \left. \opp {\tilde{\varphi}(-\alpha)}{\alpha}\right|_{\alpha=0} \!\!\!\! =  (1-\hat{\bfepsilon}_f) \cdot \widetilde{\Gamma}(0)^s \!\!\cdot (\hat{\bfepsilon}_i - \hat{\bfepsilon}_f) \,,
\end{equation}
where $A^s$ denotes the spectral (or Drazin) pseudoinverse \cite{Boullion:71}. 

\eqn{MFPT} has a simple interpretation: The steady-state ($\alpha=0$) flux  must be equal to the total number of systems in transit divided by the MFPT. Thus, when the flux equals 1 (as in this calculation), the MFPT must be equal to the sum of the occupation numbers in all the intermediate states, $\sum_{s \ne f} P_s(t)$.   \eqns{P_decomposition} and \eqnref{Pbar_solutionb} imply that the rhs of \eqn{MFPT} equals $(\bfone- \hat{\bfepsilon}_f) \cdot \bar{\bfP}$, which is identical to this sum.  We also note that \eqn{MFPT} is the same result that would be obtained if we ignore all memory effects and approximate $\Gamma(\tau) \approx \delta(\tau) \int_0^\infty \Gamma(\tau) d \tau$; this is equivalent to replacing the generalized master equation with a regular master equation having $\Gamma = \int_0^\infty \Gamma(\tau)$. Differences between the moments of these two equations only appear in higher order. 

When $\widetilde{\Gamma}(u)$ has to be determined from $\widetilde{K}(u)$, we can combine \eqns{Gamma_Laplace_relationship} and \eqnref{MFPT} [or directly \eqn{K_Laplace_relationship}] to get 
\begin{eqnarray}
\langle\langle \tau \varphi \rangle \rangle & = & \bfone \cdot \mbox{Diag}(\langle \langle \tau K \rangle \rangle) \cdot[I + \widetilde{\overline{K}}(0)]^{-1} \cdot \hat{\bfepsilon}_i \nonumber \\
& = & \bfone \cdot \mbox{Diag}(\langle \langle \tau K \rangle \rangle) \cdot \widetilde{K}(0)^s \cdot \hat{\bfepsilon}_i\,.
\label{Eqn:first_moment}
\end{eqnarray}
Thus, the MFPT is completely determined by the zeroth moments of $K(\tau)$ and the first moments of its diagonal elements, which are the mean incubation times of the individual states.  Only these $N + N_{\overline{K}}$ moments [where $N_{\overline{K}}$ is the number of non-zero off-diagonal elements in $K(\tau)$] must be determined from the molecular simulation data to compute the MFPT.  

Not only is \eqn{first_moment} simple to compute, but we expect that it will require less simulation data than direct integration to compute the MFPT to the same statistical accuracy: Integrating $d \bfP(t)/dt$ requires determination of a complete discretized representation of $K(\tau)$: i.e., the $(N+N_{\overline{K}}) \times T_\tau/h$ values of $K(n h)_{s,s'}$ for $n=0,1, \ldots T_\tau/h$, where $h$ is the smallest relevant time scale and  $T_\tau$ is the support of $K(\tau)$.  Each value must be determined by counting the number of simulation events that make the first passage from $s'$ to $s$ in time $n h \le \tau \le (n+1) h$. Since each number is a small fraction of the total number of simulation events, the statistical error of its determination will be large relative to that of the zeroth and first moments. Therefore, direct integration will require more molecular simulation data than \eqn{first_moment} to achieve the same statistical error. 

Analogous expressions for the higher FPT moments can be determined by computing additional values of $\tilde{\varphi}(-\alpha)$ near $\alpha = 0$ and numerically differentiating or by analytically expanding \eqn{relationships} or \eqnref{relationship2}  in terms of the $\langle \langle \tau^k K \rangle \rangle$ to obtain expressions analogous to \eqn{first_moment} \cite{note2}. The latter procedure can be used to compute the finite moments even when higher moments are infinite and $\tilde{\varphi}(-\alpha)$ is not analytic at $\alpha=0$.


\begin{thebibliography}{9}
\bibitem{note2} $\langle\langle \tau^k \varphi \rangle \rangle$ depends on $\mbox{Diag}(\langle\langle \tau^k K \rangle\rangle)$ and $\langle\langle \tau^{k'} K \rangle\rangle$ ($0 \le k' < k$), so only one additional set of moments of $K(\tau)$ needs to be computed for each additional FPT moment.  To maintain overall statistical accuracy, the amount of simulation data used will have to be proportionately increased.
\end{thebibliography}
\end{document}